\documentclass[showpacs,amssymb,10pt,reprint,aps,prd,longbibliography,nofootinbib,floatfix]{revtex4-2}
\usepackage{graphicx,epsfig,amssymb} 
\usepackage{amsmath,amsfonts, times}
\usepackage{bm} 
\usepackage{epstopdf}
\usepackage[linktocpage,colorlinks=true,urlcolor=blue]{hyperref}
\usepackage{subfigure}
\usepackage[usenames,dvipsnames]{xcolor}     
\usepackage{natbib}
\usepackage{soul}
\usepackage[utf8x]{inputenc}
\usepackage{float}
\definecolor{darkgreen}{rgb}{0,0.5,0}

\def\p{\partial}
\def\bh{\text{BH}}

\DeclareUnicodeCharacter{2009}{\,} 

\begin{document}
\title{Shadows of black holes with dark matter halo}

	\author{S\'ergio V. M. C. B. Xavier}
			\email{sergio.xavier@icen.ufpa.br}
			\author{Haroldo C. D. Lima Junior}
			\email{haroldolima@ufpa.br}
	\author{Lu\'is C. B. Crispino}
		\email{crispino@ufpa.br}
	\affiliation{Programa de P\'os-Gradua\c{c}\~{a}o em F\'{\i}sica, Universidade 
		Federal do Par\'a, 66075-110, Bel\'em, Par\'a, Brazil.}

\begin{abstract}
We investigate the shadow of an exact black hole solution of Einstein's equations recently proposed by Cardoso {\it et~al.}, to describe a supermassive black hole immersed in a dark matter halo. We analyze and discuss the light rings and the gravitational lensing of this spacetime comparing them with an isolated Schwarzschild black hole. Using backward ray-tracing techniques, we study the shadows cast by such black hole when illuminated by a celestial sphere that emits radiation isotropically. We find that when the dark matter distribution concentrates near the event horizon of the black hole, multiple light rings emerge. In this high compactness regime, the shadows and gravitational lensing are significantly different from the Schwarzschild one. We also use the M87* and SgrA* shadow data, obtained by the Event Horizon Telescope collaboration, to constrain the parameters of the dark matter halo.
	\end{abstract}
	
	\date{\today}
	
	\maketitle

\section{Introduction}\label{sec:int}
	The groundbreaking results of the Event Horizon Telescope (EHT) collaboration powered the field of black hole (BH) imaging research. The image of the M87* BH, released in 2019~\cite{M87_1:2019,M87_2:2019,M87_3:2019,M87_4:2019,M87_5:2019,M87_6:2019}, inspired  several studies on BH shadows to test different gravity theories and to investigate features of the electromagnetic resemblance of BHs (for some recent works see, {\it e.g.}, Refs.~\cite{Cunha_etal:2015,Abdujabbarov_etal:2016,Cunha_etal:2016,Cunha_etal:2017PRL,Tsukamoto:2018,Cunha_etal:2019,Badia_Eiroa:2020, Guerrero_etal:2021,Adrin_etal:2021,Volkel_etal:2021,Haroldo_etal:2021}). The recent EHT results, released in May 2022, targeted the center of our galaxy and revealed the image of SgrA* surrounded by an asymmetric emission ring~\cite{sgra_1:2022,sgra_2:2022,sgra_3:2022,sgra_4:2022,sgra_5:2022,sgra_6:2022}. These recent results increased the statistical significance of black hole shadow data, which can be used to constrain even further alternative black hole models beyond the Kerr hypothesis \cite{Herdeiro:2022}.
	
	The edge of a BH shadow is determined by light rays that approach asymptotically bound photon orbits~\cite{Gralla_etal:2019,Bronzwaer_Falcke:2021}. In the case of non-spherical spacetimes, these orbits are called fundamental photon orbits (FPOs) \cite{Cunha_etal:2017}, generalizing the planar light rings (LRs), which fully describe the shadows in spherically symmetric spacetimes. For the particular case of Kerr spacetime, where the Hamilton-Jacobi equation is separable in the Boyer-Lindquist coordinates, all the FPOs are known as spherical photon orbits (SPO) \cite{Teo:2003}.
	
It is unlikely that BHs are completely isolated objects in the universe. The astrophysical scenarios where most of these objects live are rich environments.~In particular, there is strong evidence that supermassive BHs source active galactic nuclei~\cite{Rees:1984, Kormendy_Richstone:1995,Richstone:1998}. On these galactic stages, another character plays an important role: dark matter. There is a large amount of evidence that dark matter surrounds most galaxies in a halo extending beyond the luminous part of the galaxy~\cite{Bertone_Tait:2018}. Dark matter constitutes one of the most important puzzles in modern astronomy. Some studies about the implications of dark matter in the image of BHs were performed in the last years (see Refs.~\cite{Hou_etal:2018,Jusufi_etal:2019,Konoplya:2019,Saurabh_Jusufi:2021,Nampalliwar_etal:2021,Qian_etal:2022} for some recent works and references therein).

Recently, Cardoso {\it et al.} proposed an exact solution of Einstein gravity minimally coupled with an anisotropic fluid, representing a BH immersed in a dark matter halo described by a Hernquist profile~\cite{Cardoso_etal:2021}. In that paper, the authors studied the influence of the dark matter component in the propagation of gravitational waves (GW) and geodesics (at the linear level). In Ref.~\cite{Konoplya:2021}, the electromagnetic perturbations and the Unruh temperature were studied within that dark matter configuration, and the authors of Ref.~\cite{Stuchlik_Vrba:2021} analyzed the epicyclic oscillatory motion of test particles. Tidal forces generated by this BH-dark matter system were investigated in Refs.~\cite{Junior_etal:2022, Liu_etal:2022}. Other BH models with different dark matter halo distributions were obtained in Refs.~\cite{Jusufi:2022,Konoplya_Zhidenko:2022}.

We extend the geodesic analysis made at the linear level in Ref.~\cite{Cardoso_etal:2021}. In particular, we investigate the shadows and gravitational lensing, using backward ray-tracing methods, for a wide range of the halo compactness parameter. Moreover, we use the EHT data of M87* and SgrA* to constrain the parameters of the dark matter halo of the solution proposed in Ref.~\cite{Cardoso_etal:2021}.

The remainder of this paper is organized as follows. In Sec.~\ref{sec:2}, we review the properties of the spacetime of a BH immersed in a Hernquist distribution of matter.  In Sec.~\ref{sec:3}, we study the motion of null geodesics and the LRs. In Sec.~\ref{sec:4}, we discuss the shadow and lensing of such BH when illuminated by a celestial sphere. Using the EHT observations of M87* and SgrA*, in Sec.~\ref{sec:5} we report constraints on the parameters of this metric. Our final remarks are presented in Sec.~\ref{sec:remarks}. Unless otherwise stated, natural units $c = G = 1$ are used.

\section{The geometry of asymptotically flat black holes with hair and regular horizon} \label{sec:2}

In order to construct an exact field solution for a BH surrounded by dark matter, the authors of Ref.~\cite{Cardoso_etal:2021} employed a generalized version of the ``Einstein cluster" -- a technique that generates a stationary system of many gravitating masses. To model the galactic profiles, they used a Hernquist-type density distribution, namely
	\begin{equation}
		\rho(r)=\frac{Ma_{0}}{2\pi r (r+a_{0})^{3}},
		\label{Hernquist}
	\end{equation}
where $M$ is the total mass of the halo and $a_{0}$ is a typical length-scale of the galaxy.

 The geometry of the resulting spacetime is described by the following line element:
	\begin{equation}
		ds^{2}=-f(r)dt^{2}+\frac{dr^{2}}{1-2m(r)/r}+r^2d\Omega^2 \, ,
		\label{Cardosoetalmetric}
	\end{equation}
where $d\Omega^{2}$ is the line element of a $2$-sphere of unitary radius. The function $f(r)$ is obtained from the mass function $m(r)$, which, for this spacetime, is given by:
	\begin{equation}
		m(r)=M_{\bh}+\frac{M r^2}{(a_{0}+r)^2}\left(1-\frac{2M_{\bh}}{r}\right)^2\,.
	\end{equation}
In large scales, this mass function corresponds to a distribution of matter described by the Hernquist profile \eqref{Hernquist} and, at small distances, it describes a BH with mass $M_{\bh}$.

After imposing the asymptotic flatness condition, one obtains an analytic solution of Einstein's equations corresponding to the following function $f(r)$:
	\begin{equation}
		f(r)=\left(1-\frac{2M_{\bh}}{r}\right)e^{\Upsilon},
	\end{equation}		
	with
	\begin{align}
		&\Upsilon=-\pi\sqrt{\frac{M}{\xi}}+2\sqrt{\frac{M}{\xi}}\arctan\left(\frac{r+a_{0}-M}{\sqrt{M\xi}}\right),\\
		&\xi=2a_{0}-M+4M_{\bh}.
	\end{align}
The matter density associated with this profile is 
	\begin{equation}
		4\pi\rho(r)=\frac{2M(a_{0}+2 M_{\bh})(1-2M_{\text{BH}}/r)}{r(r+a_{0})^3}.
		\label{matterdensity}
	\end{equation}
Note that the event horizon is located at $r_{\text{H}}=2M_{\bh}$, as in the Schwarzschild solution. The ADM mass of this spacetime is the sum of the mass of the BH and the mass of the halo of matter, $M_{\text{ADM}}= M+M_{\text{BH}}$.

We will focus mainly on the astrophysical setup that is determined by the regime:
	\begin{equation}
		\label{astro_setup}M_{\text{BH}}\ll M \ll a_{0},
	\end{equation}
although we also study some particular cases beyond this range.
To compare different halo configurations, one can define the compactness parameter as:
	\begin{equation}
	\mathcal{C}=\frac{M}{a_0}.
	\end{equation}
In galactic contexts, this parameter can assume values greater than $10^{-4}$ \cite{Cardoso_etal:2021}.

\section{Null geodesics and Light rings} \label{sec:3} 
Null geodesics can be obtained by Hamilton's equations:
	\begin{align}
		&\dot{x}^{\mu}=\frac{\partial\mathcal{H}}{\p p_{\mu}}\label{xdot},\\
		&\dot{p}_{\mu}=-\frac{\partial\mathcal{H}}{\p x^{\mu}},
	\end{align}
where $\mathcal{H}$ is the Hamiltonian, given by
	\begin{equation}
		\mathcal{H}=\frac{1}{2}g^{\mu\nu}p_{\mu}p_{\nu}=0.
		\label{Hphoton}
	\end{equation}

The spherical symmetry allows us to perform the analysis in the equatorial plane ($\theta=\pi/2$) without loss of generality. 
Besides that, the axial symmetry, together with stationarity, guarantees that this spacetime has the following constants along the geodesic motion: 
	\begin{align}
		&p_{t}=-f(r)\dot{t}=-E,\label{teq}\\
		&p_{\varphi}=L,
	\label{phieq}
	\end{align}
where $E$ and $L$ are the energy and the total angular momentum of the photon, respectively. 

Following Ref.~\cite{Cunha_etal:2017PRL}, we can write the Hamiltonian in the form
	\begin{equation}
		\mathcal{H}=T(r,\theta)+V(r,\theta, E, L) \, ,
	\end{equation}
where $T$ and $V$ are kinetic and potential terms, respectively, defined as:
	\begin{align}
		&T\equiv p_{r}^{2}g^{rr}+p_{\theta}^{2}g^{\theta\theta},\\
		&V\equiv p_{t}^{2}g^{tt}+p_{\varphi}^{2}g^{\varphi\varphi}.
	\end{align}
In order to perform an analysis independent of $E$ and $L$, we can define the effective potential $H(r)$, related to $V$, as
	\begin{equation}
		V(r, \pi/2, E, L)=-\frac{L^2}{f(r)}\left(\frac{1}{b}-H(r)\right)\left(\frac{1}{b}+H(r)\right),
	\end{equation}
where $b\equiv L/E$ is the impact parameter. The effective potential $H(r)$ can be written as: 
	\begin{equation}
		H(r)=\frac{\sqrt{f(r)}}{r}.
		\label{effecpotential}
	\end{equation}
A LR, in this approach, corresponds to a critical point of $H(r)$, i.e.,
	\begin{equation}
		\label{Eq-LR}H'(r_{\text{LR}})=0 \, ,
	\end{equation}
with the prime denoting derivative with respect to the radial coordinate. The impact parameter related to the LR is given by
	\begin{equation}
		\label{crit-parameter}b_{\text{crit}}=\frac{1}{H(r_{\text{LR}})}.
	\end{equation}
As shown in Ref.~\cite{Cardoso_etal:2021}, keeping only terms up to $\mathcal{O}(1/a^3_{0})$, the LR, in this geometry, lies at
	\begin{equation}
		r_{\text{LR}}\approx 3M_{\bh}\left(1+\frac{MM_{\bh}}{a_{0}^{2}}\right),
	\end{equation}
and the critical impact parameter $b_{\text{crit}}$ is given by:
	\begin{equation}
		b_{\text{crit}}=3\sqrt{3}M_{\bh}\left(1+\frac{M}{a_{0}}+\frac{M(5M-18M_{\bh})}{6a_{0}^{2}}\right).
	\end{equation}

\begin{figure}[h!]
\includegraphics[width=8cm]{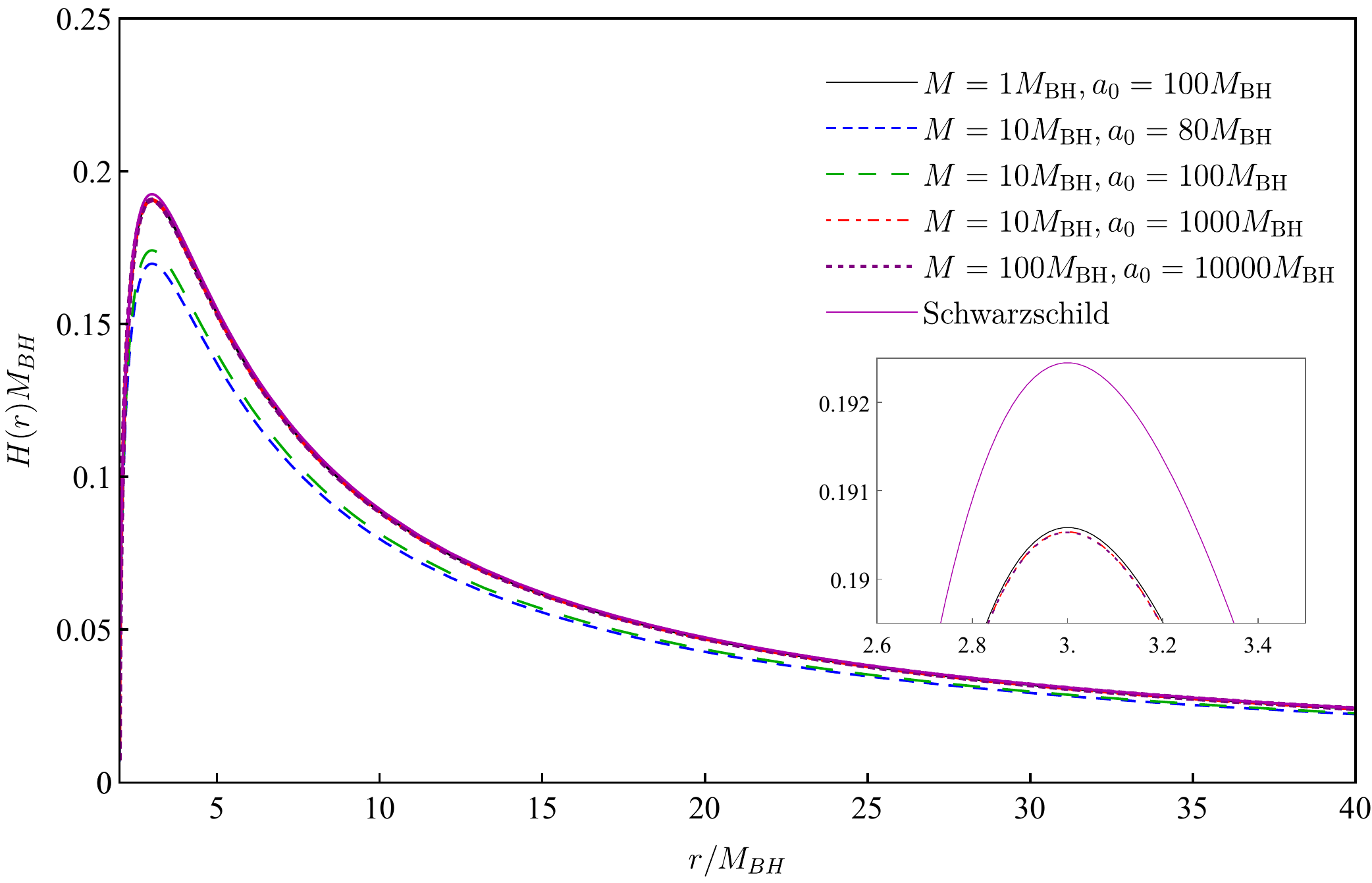}\vspace{0.2cm}
\includegraphics[width=8cm]{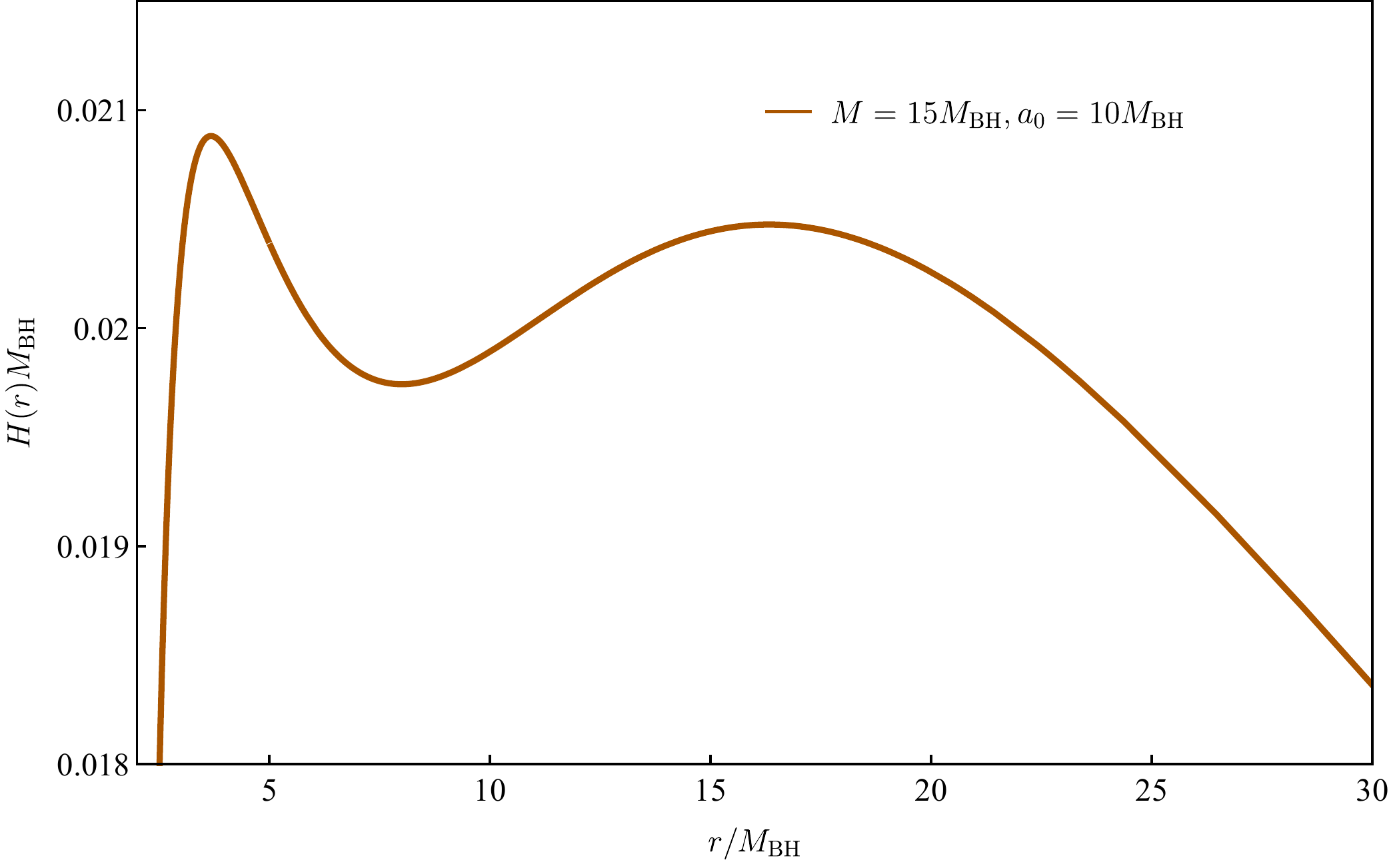}
\caption{\label{figeffecpotential} Effective potential $H(r)$ of the spacetime represented by Eq.~\eqref{Cardosoetalmetric} for different values of the halo properties, namely $\mathcal{C}\ll 1$ (top panel) and $\mathcal{C} = 1.5$ (bottom panel).}
\end{figure}
\begin{figure}
\includegraphics[width=8cm]{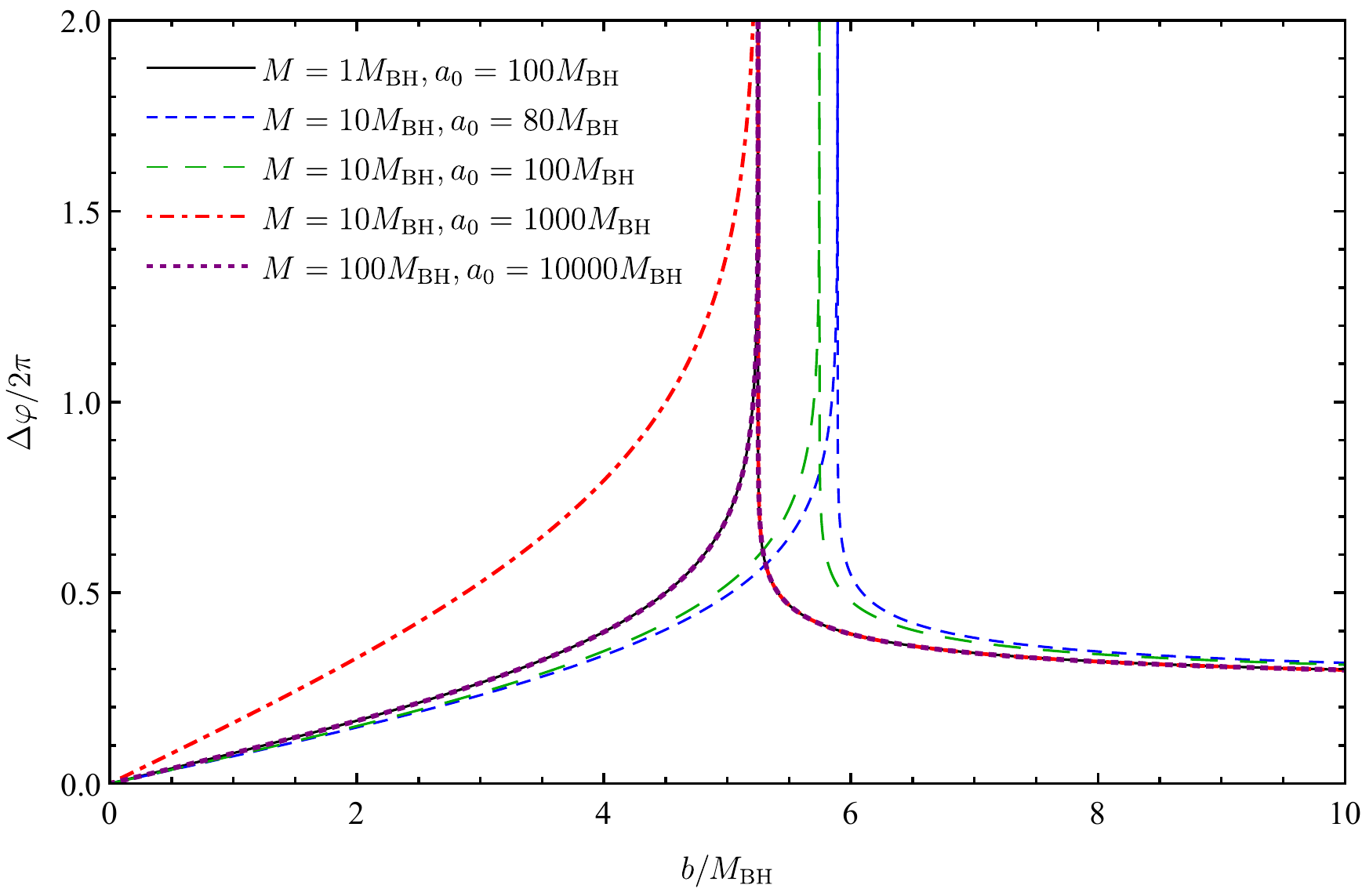}\vspace{0.2cm}
\includegraphics[width=8cm]{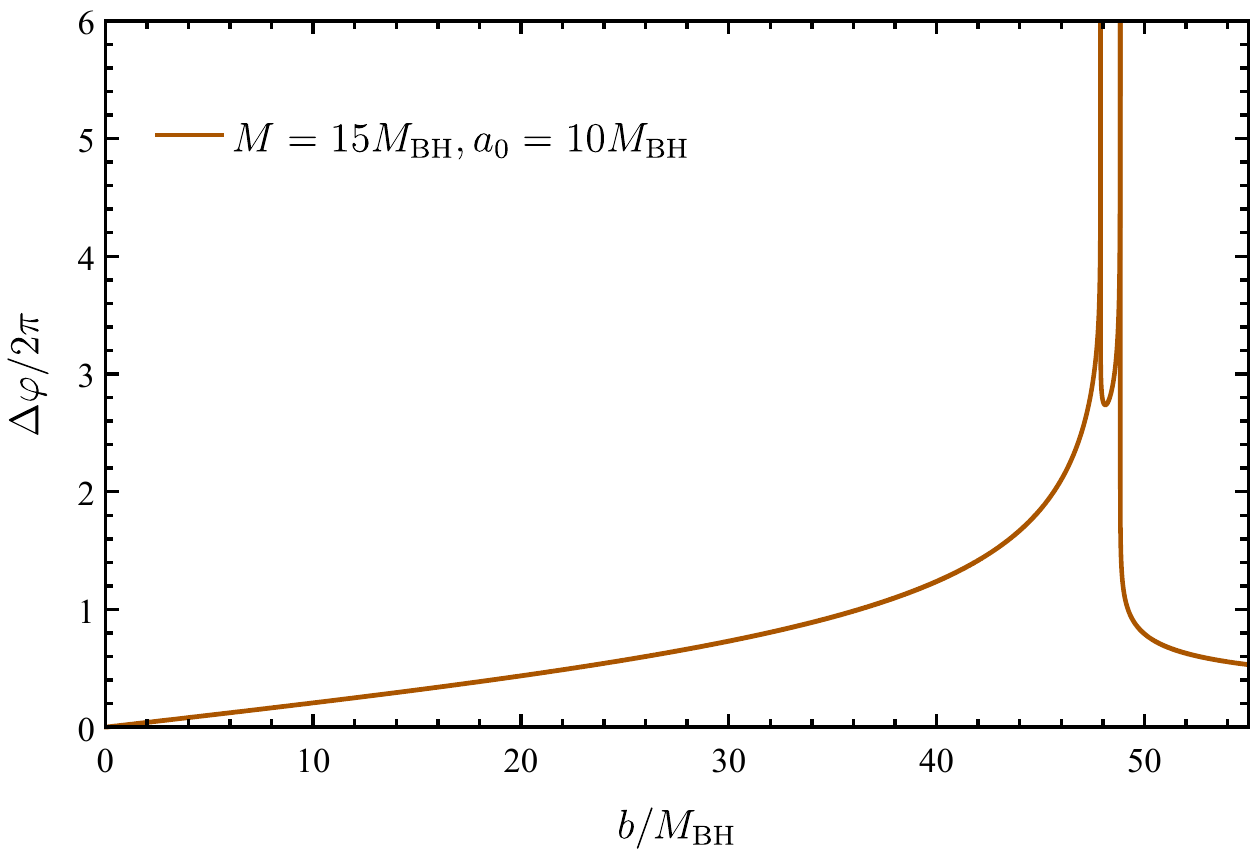}\vspace{0.2cm}
\includegraphics[width=8.5cm]{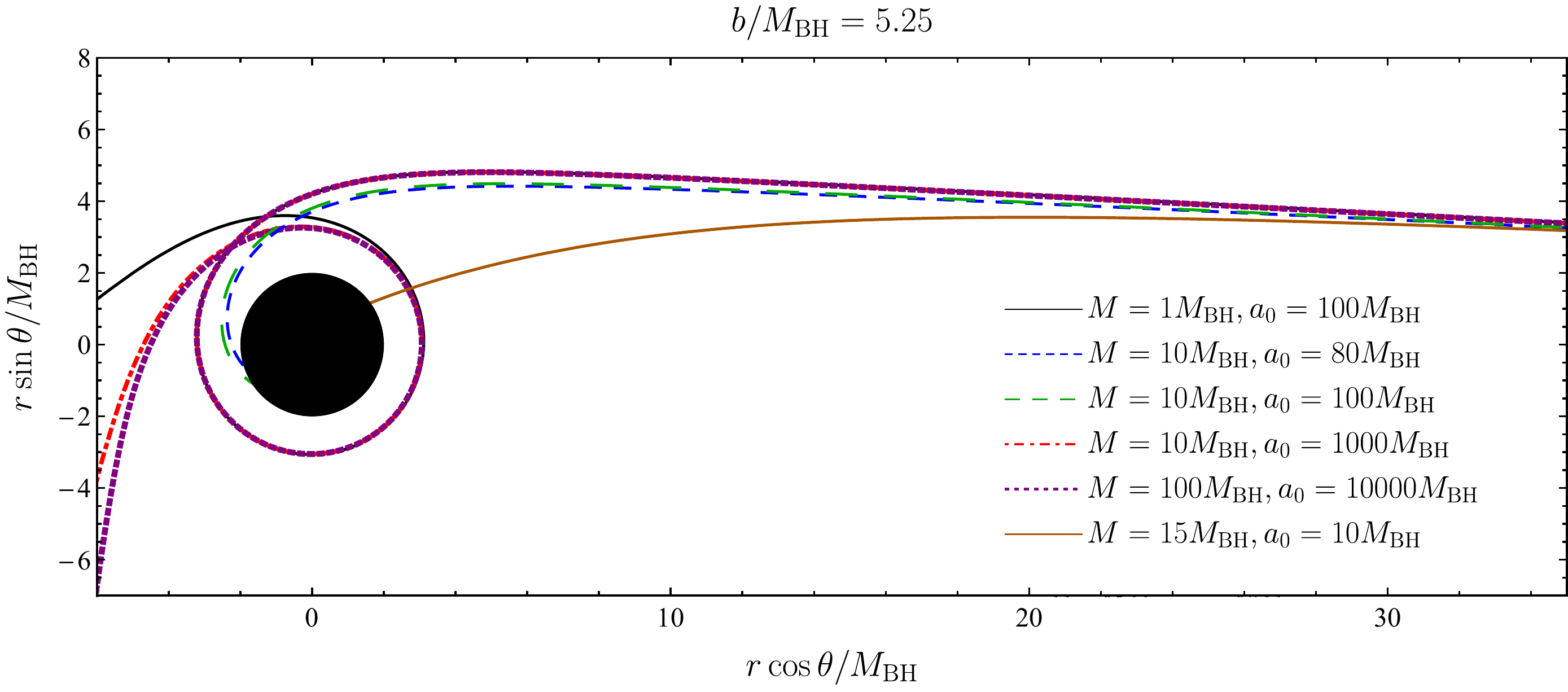}\vspace{0.2cm}
\caption{{\it Top panel}: Scattering angle for null geodesic, as a function of the impact parameter, for different values of $a_{0}/M_{\bh}$ and $M/M_{\bh}$. {\it Middle panel:} Scattering angle for the case with $\mathcal{C}= 1.5$ ($M = 15M_{\bh}$ and $a_0=10M_{\bh}$). {\it Bottom panel}: Null geodesics with impact parameter $b/M_{\bh}=5.25$ and different values of the halo parameters, including the case $\mathcal{C} = 1.5$.}
\label{scattangle}
\end{figure}

In the top panel of Fig.~\ref{figeffecpotential}, we show the effective potential $H(r)$, given by Eq.~\eqref{effecpotential}, for different values of $M/M_{BH}$ and $a_0/M_{BH}$ corresponding to small values of the compactness parameter $\mathcal{C}$. In this regime, one can see the local extremum associated with an unstable LR. In the bottom panel of Fig.~\ref{figeffecpotential}, we exhibit an example of a halo configuration with compactness $\mathcal{C} = 1.5$ ($M = 15M_{\bh}$ and $a_0=10M_{\bh}$). {Although this case may not describe plausible astrophysical BH-halo systems, it presents features of theoretical interest. In such case, one can identify three local extrema corresponding to two unstable LRs (represented by the two local maxima of the effective potential) and one stable LR -- characterized by the presence of a local minimum in $H(r)$ -- in agreement with the topological charge conservation results of Ref.~\cite{Cunha_Herdeiro:2020}.

From Eqs.~\eqref{Hphoton}, \eqref{teq} and \eqref{phieq}, one can obtain the following expression:
	\begin{equation}
		\label{Eqshape}
		\frac{f(r)}{r^{2}}\left[\frac{1}{r^2\left(1-2m(r)/r\right)}\left(\frac{dr}{d\varphi}\right)^{2}+1\right]-\frac{1}{b^2}=0.
	\end{equation}
After integrating Eq.~\eqref{Eqshape}, one finds the total variation in the $\varphi$ angle. In Fig.~\ref{scattangle}, we exhibit the scattering angle as a function of the impact parameter. For values of $b$ close to $b_{\text{crit}}$, we note a divergence in $\Delta\varphi$, what means that the photon winds several turns around the BH before being scattered. For the case when there are multiple LRs ($\mathcal{C} = 1.5$), the scattering angle presents more than one divergent position associated with each unstable LR. 
In Fig.~\ref{scattangle} we also exhibit null trajectories at the equatorial plane for the same value of the impact parameter $b/M_\bh$ and different values of $M/M_\bh$ and $a_0/M_\bh$. One can see that,  by increasing the compactness, for a fixed impact parameter, light rays that would initially escape to infinity are trapped in the event horizon. This behavior is due to an enormous concentration of dark matter around the BH, which causes the critical impact parameter to increase. 
This property is evident, in particular, for the case $\mathcal{C} = 1.5$, shown in the bottom panels of Figs.~\ref{figeffecpotential} and~\ref{scattangle}.

\section{Black hole shadow and gravitational lensing} \label{sec:4}
\begin{figure}[h!]
\includegraphics[width=8cm]{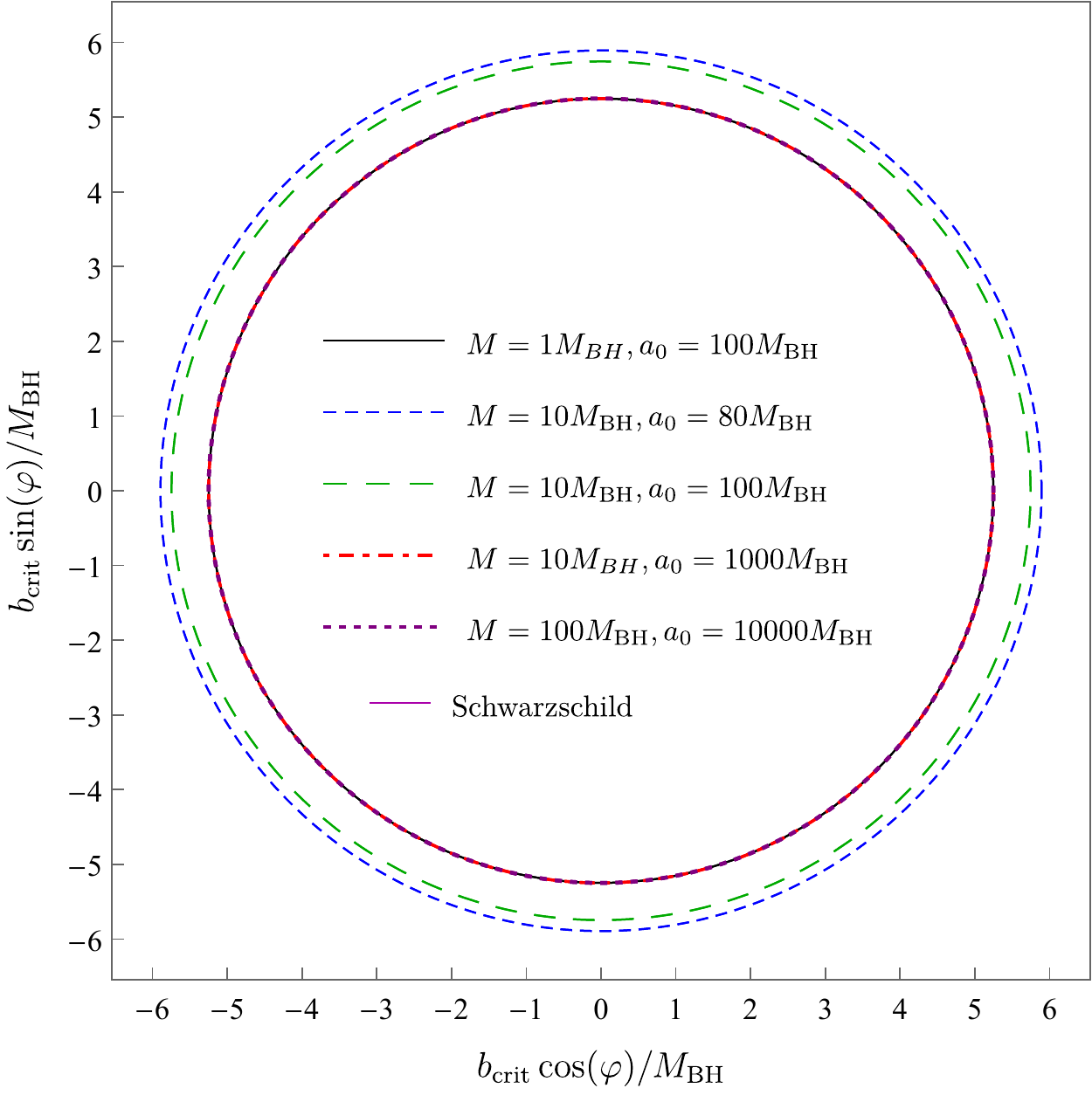}\vspace{0.2cm}
\includegraphics[width=8cm]{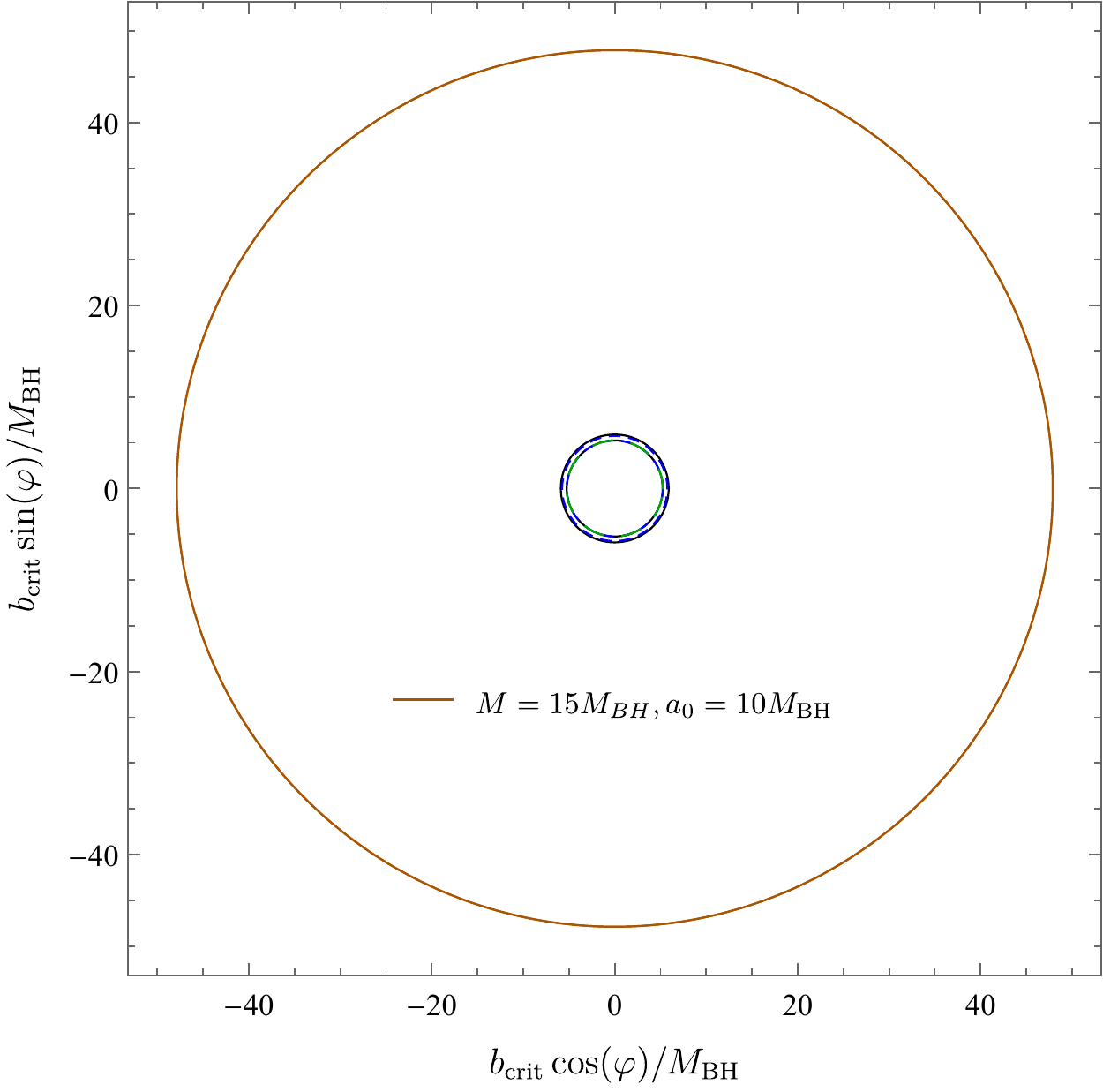}
\caption{{\it Top panel}: Shadow edge of the BH solution~\eqref{Cardosoetalmetric} for different values of $M$ and $a_{0}$, in the low compactness regime. {\it Bottom panel}: Shadow edge for the case with $\mathcal{C} = 1.5$ (high compactness), compared with the low compactness regime.}
\label{shadow}
\end{figure}
\begin{figure}
\centering
\subfigure[\ Schwarzschild]{\includegraphics[width=6.5cm]{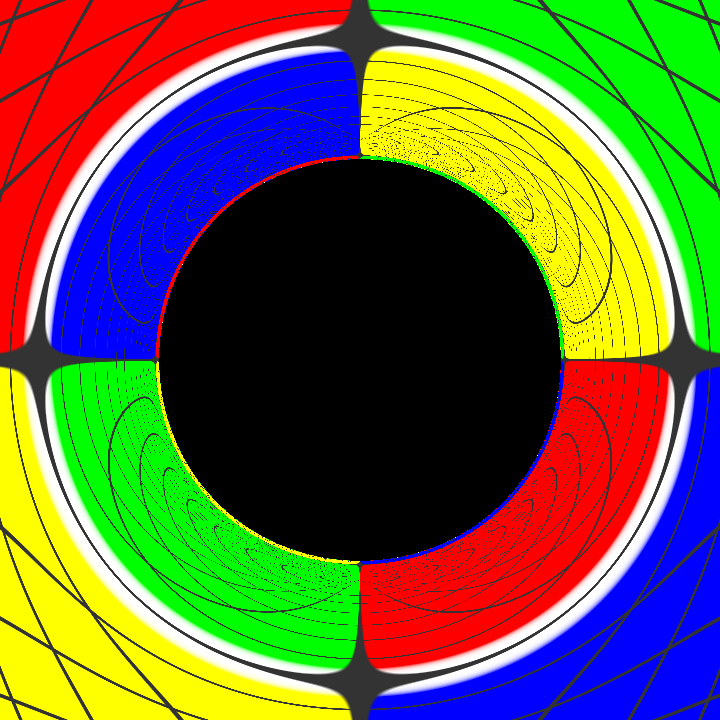}}
\subfigure[\ $M=10M_{\bh}$; $a_0=1000M$]{\includegraphics[width=6.5cm]{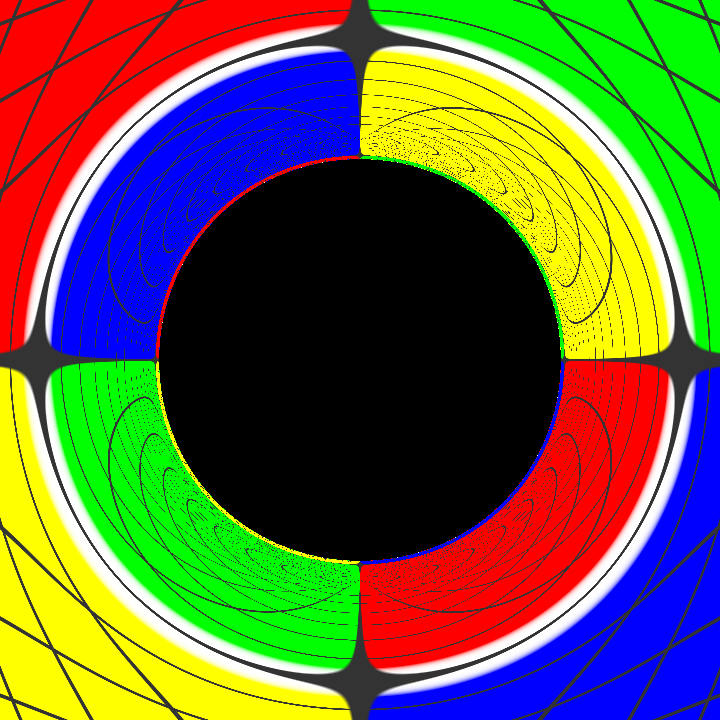}}
\subfigure[\ $M=15M_{\bh}$; $a_0=10M_{\bh}$]{\includegraphics[width=6.5cm]{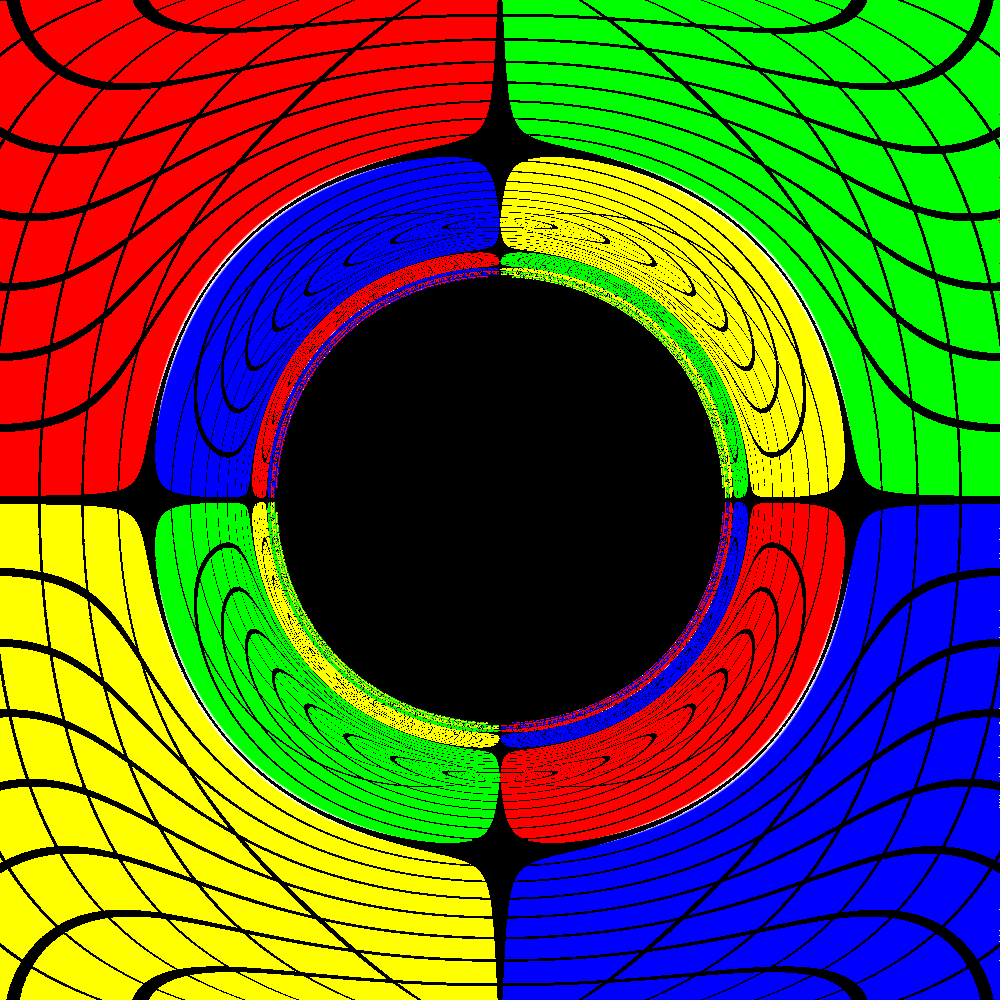}}
\caption{Shadow and gravitational lensing of the spacetime solution~\eqref{Cardosoetalmetric}. In panels (a) and (b), the observer is located at the equatorial plane with radial coordinate $R_{\text{obs}} = 15M_{BH}$. In the panel (c), the observer is located far away from the event horizon.}
\label{Figlensing} 
\end{figure}

In this section, we investigate the shadow and the gravitational lensing of the BH solution \eqref{Cardosoetalmetric} when illuminated by a celestial sphere, which is concentric with the BH and emits radiation isotropically. The boundary of the BH shadow, in these spherically symmetric cases, is related to the critical impact parameter associated with the LR \cite{Gralla_etal:2019}. In Fig.~\ref{shadow}, we show the shadow edge of the BH solution \eqref{Cardosoetalmetric}, as seen by an observer at infinity, for different choices of $M$ and $a_{0}$, compared to an isolated Schwarzschild BH.

As mentioned in Ref.~\cite{Cardoso_etal:2021}, at dominant order, the effect of the mass halo is essentially to redshift the dynamics, which means that there is no deformation of the spherical symmetry of the shadow.  Therefore, a Schwarzschild BH and the BH solution \eqref{Cardosoetalmetric} with the same ADM mass will have a similar shadow. One can show that the shadow degeneracy conditions found in Ref.~\cite{Haroldo_etal:2021} are not satisfied, which means that the shadow of solution \eqref{Cardosoetalmetric} is not degenerate with the Schwarzschild shadow. Moreover, it is worth drawing attention to the fact that although the shadow sizes are not the same, it is unlikely that the EHT observations would perceive this difference, as we discuss in Sec. \ref{sec:5}. 

In the bottom panel of Fig.~\ref{shadow}, we show the shadow edge for the case with high compactness, together with the low compactness cases. We notice that the shadow size for the high compactness case is much larger than the low compactness ones.

In Fig.~\ref{Figlensing}, we show the shadow and gravitational lensing of the BH solution \eqref{Cardosoetalmetric} in comparison with the Schwarzschild case. These images were obtained using the backward ray-tracing technique, which consists in numerically evolving, backward in time, the null geodesics from the observer position until their absorption by the event horizon or scattering to infinity. The setup used is the same as in Refs.~\cite{Cunha_etal:2015, Bohn_etal:2015}, where we divided the celestial sphere (which is the light source)  into four quadrants with different colors: red, green, blue and yellow (RGBY). For each light ray that falls into the event horizon, we assign a dark pixel, whereas for light trajectories scattered to infinity, we designate one of the four colors of the grid. These images were generated by the  employment of the  \texttt{PYHOLE} code \cite{Cunha_etal:2016}. We also cross-checked our results with a C\texttt{++} code.

For the configuration shown in the middle panel of Fig.~\ref{Figlensing}, the shadow and gravitational lensing are very similar to the Schwarzschild case (top panel of Fig.~\ref{Figlensing}), although not equal. When there is a concentrated dark matter distribution near the event horizon of the BH (high values of $\mathcal{C}$), the shadow and lensing pattern is significantly different from the Schwarzschild one. In the bottom panel of Fig.~\ref{Figlensing} we show the shadow and gravitational lensing for a BH geometry with $M=15M_{\bh}$ and $a_0=10M_{\bh}$ ($\mathcal{C} = 1.5$). These values correspond to the same configuration considered in the bottom panels of Fig.~\ref{figeffecpotential}, ~\ref{scattangle} and~\ref{shadow}. We note that this BH geometry is not in the astrophysical regime specified in Eq.~\eqref{astro_setup}. However, it presents several interesting properties from a theoretical point of view. For this configuration, we have chosen the observer to be outside the LRs and far away from the event horizon. We note that this BH geometry possesses three LRs: two unstable and one stable. The existence of two unstable LRs gives rise to a gravitational lensing quite different from the Schwarzschild one, as can be seen in the bottom panel of Fig.~\ref{Figlensing}. We also note that the shadow radius is much larger than the Schwarzschild value with the same $M_{\bh}$, which is why we have chosen the radial coordinate of the observer to be greater than the other cases in Fig~\ref{Figlensing}. The shadow boundary is determined by the unstable LR associated with the highest potential barrier, the so-called \textit{dominant} LR~\cite{Haroldo_etal:2021}. From the bottom panel of Fig.~\ref{figeffecpotential}, we notice that, for $M=15M_{BH}$ and $a_0=10M_{BH}$, the dominant LR is the one closest to the event horizon.

\section{Constraints from the EHT observations of M87* and SgrA*}\label{sec:5}
%
\begin{figure*}
\includegraphics[width=8cm]{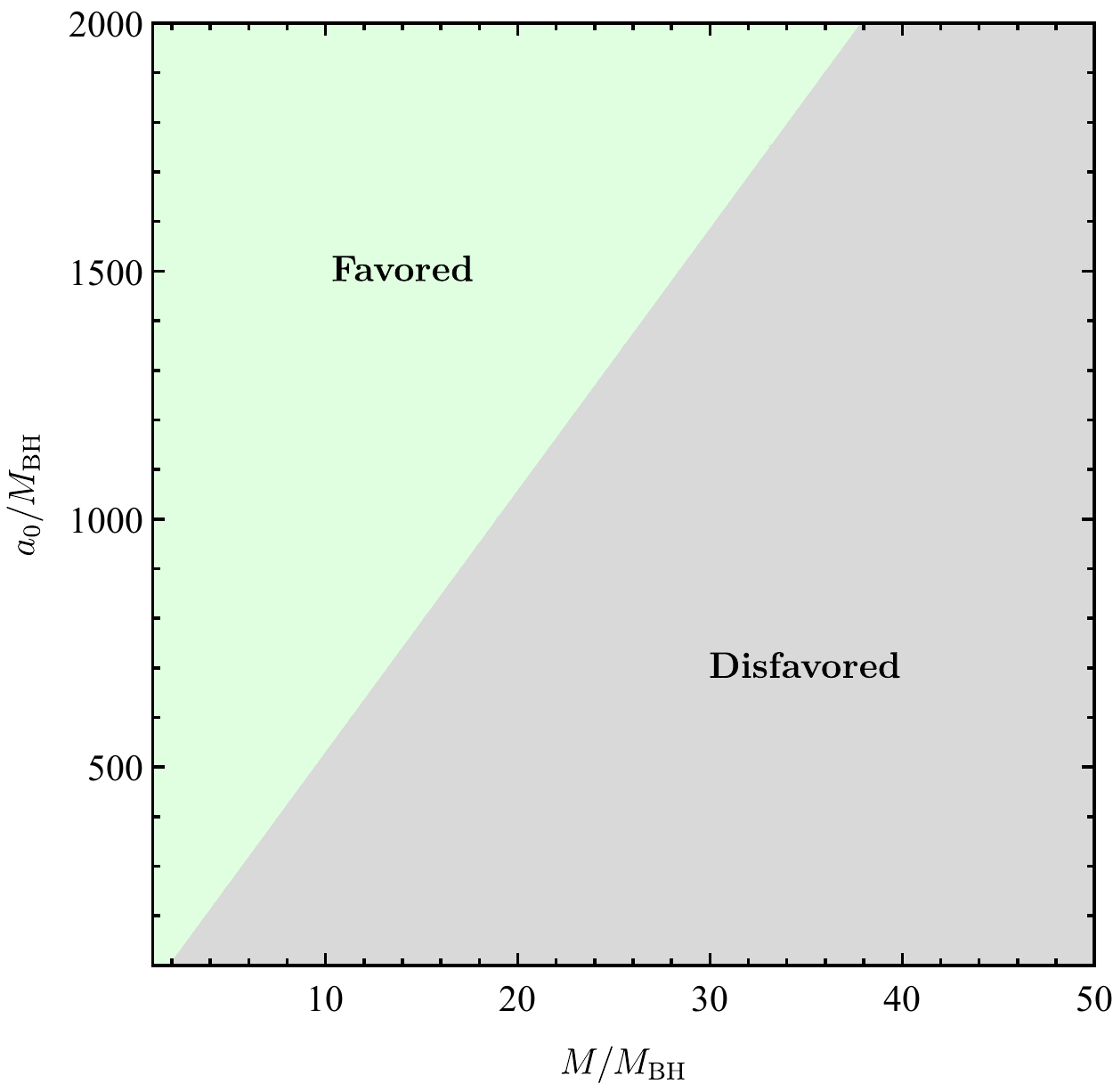}
\includegraphics[width=8.3cm]{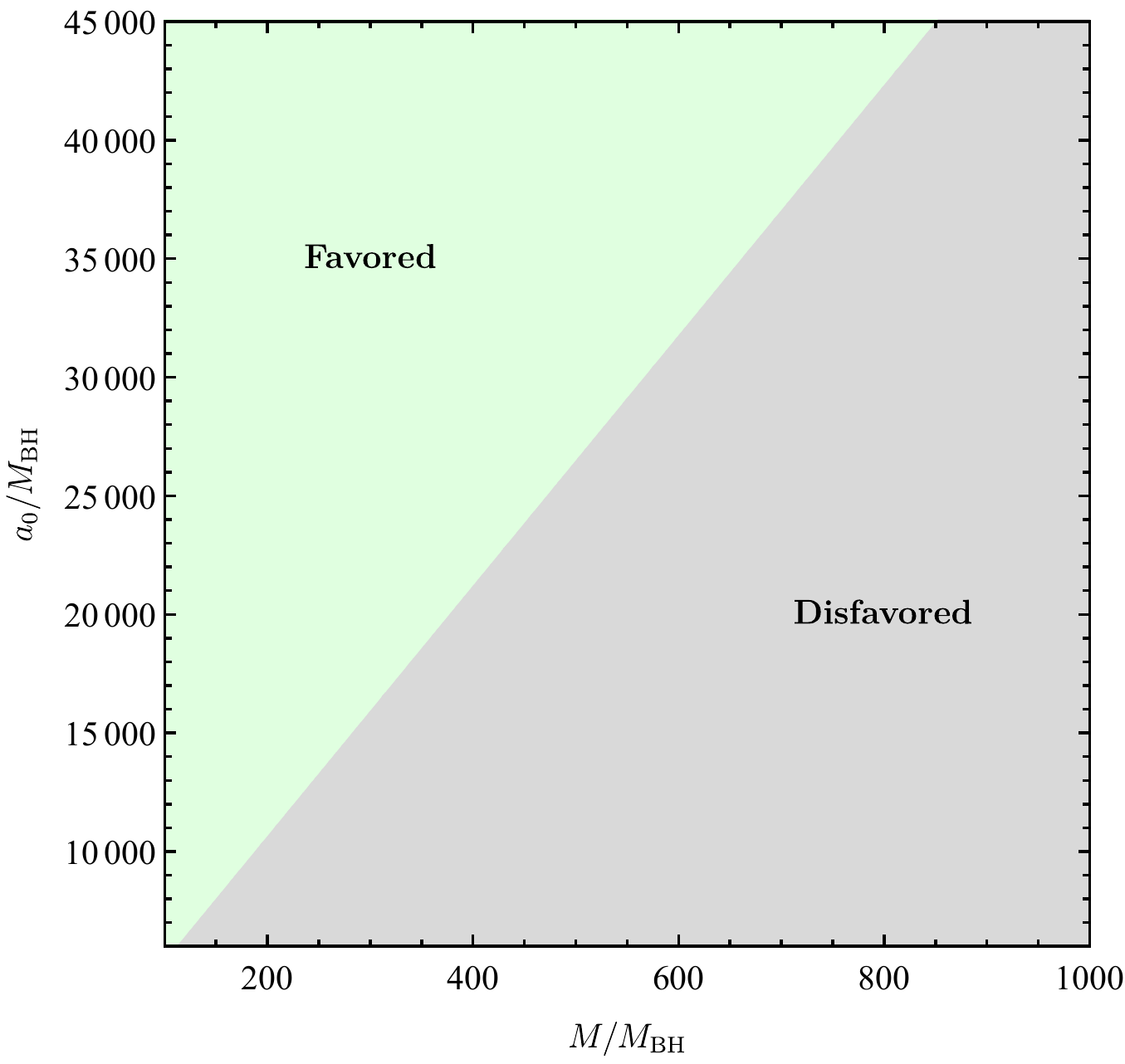}
\caption{Constraint for the geometry~\eqref{Cardosoetalmetric} on the ($M,a_0$)-plane, based on the EHT results for setups with $M_\bh< M$ (left) and $M_\bh\ll M$ (right). 
In the favored (green/lighter) region the shadow of the BH geometry~\eqref{Cardosoetalmetric} deviates less than $10\%$ from the Schwarzschild result, whereas in the unfavored (gray/~darker) region, the deviation is more than $10\%$.}
\label{EHT-constraint}
\end{figure*}
\begin{figure*}
\includegraphics[width=8.24cm]{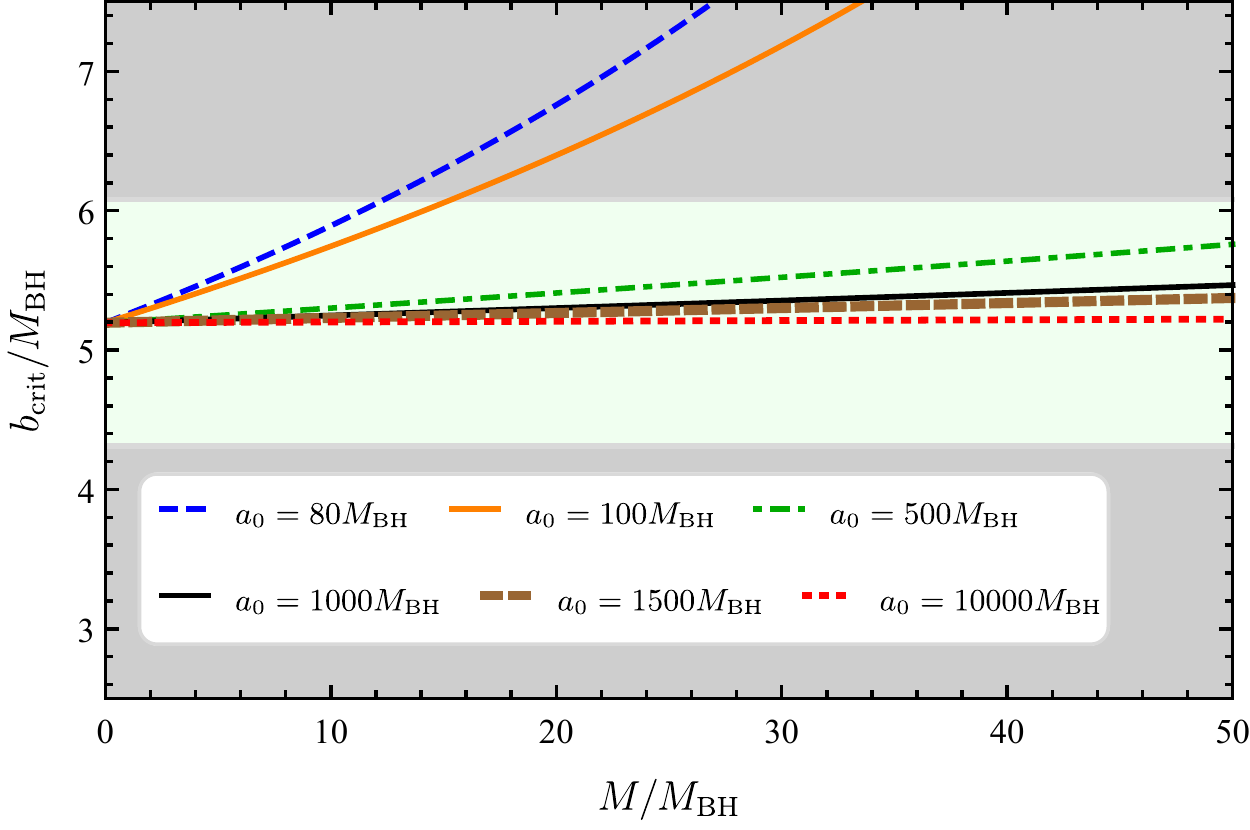}
\includegraphics[width=8.24cm]{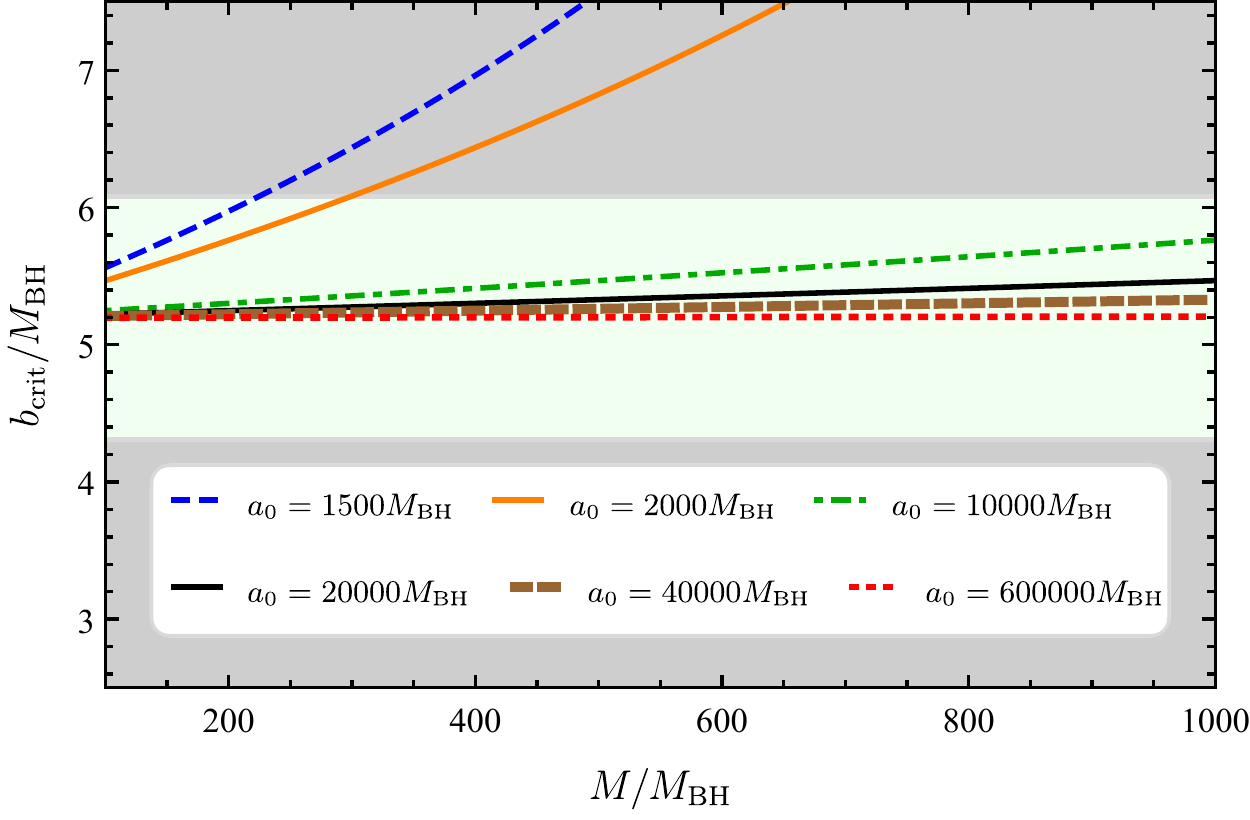}\\
\includegraphics[width=8.24cm]{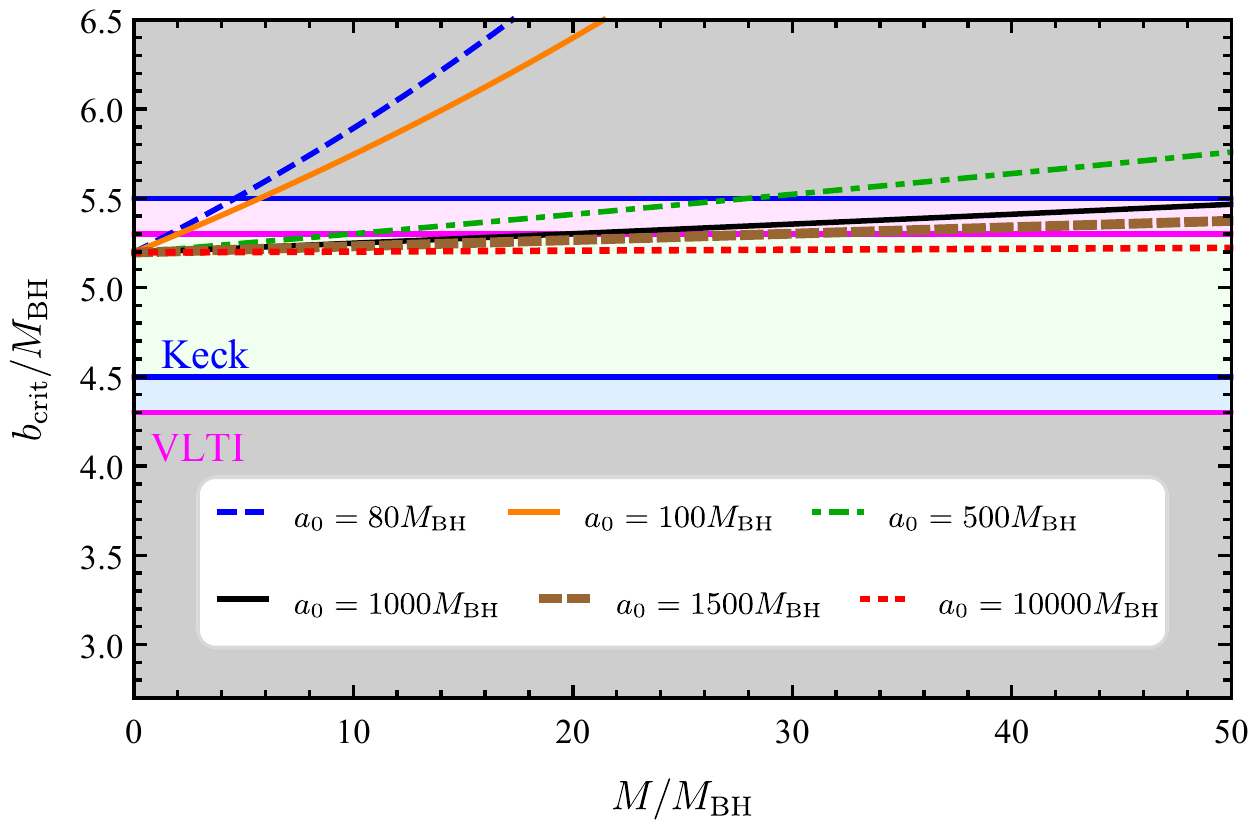}
\includegraphics[width=8.24cm]{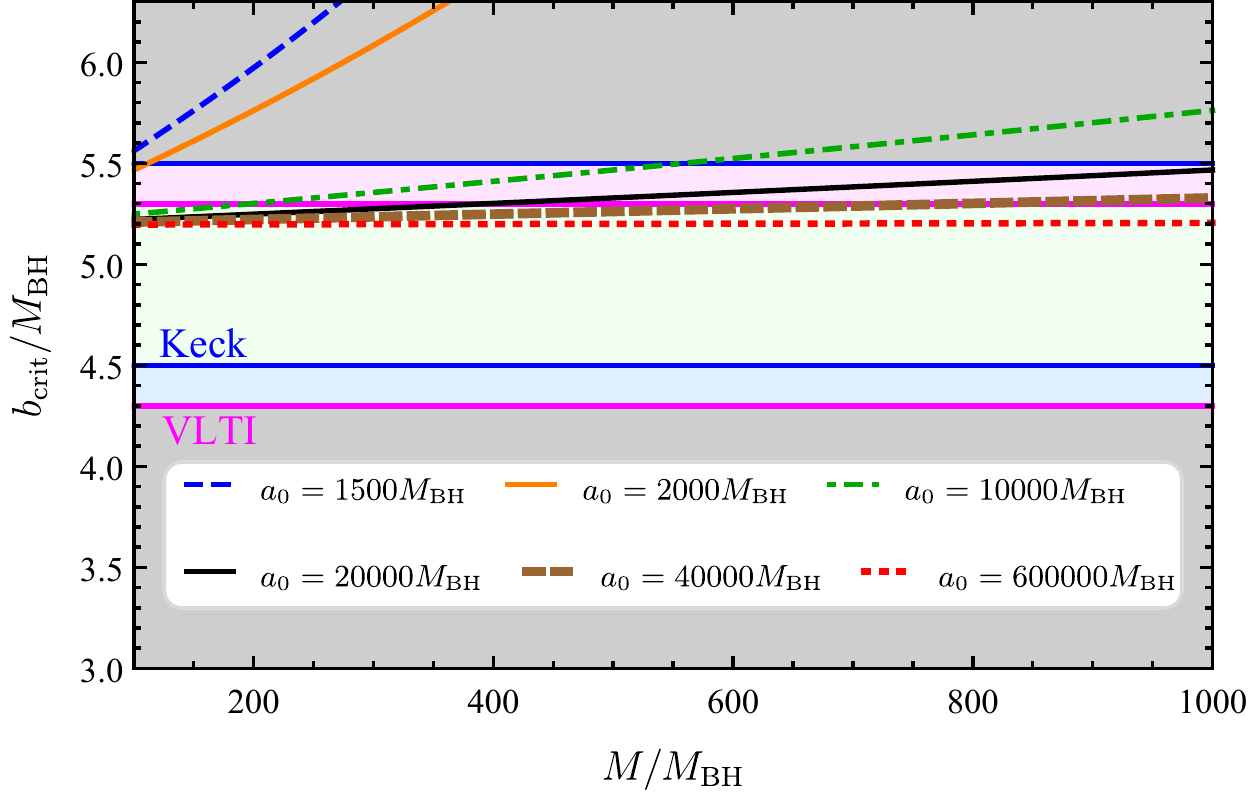}
\caption{Shadow radius as a function of $M$ for different values of $a_{0}$. The gray (darker) regions correspond to values of $M$ and $a_{0}$ inconsistent with the observations of stellar dynamics for M87* (top row) and SgrA* (bottom row). In the top row, the green (lighter) regions denote configurations consistent with the stellar dynamics for M87*. In the bottom row, the blue and pink regions denote configurations compatible with the Keck and VLTI observations, respectively. The green region is the intersection of the two compatible regions (pink and blue). The plots on the left column correspond to configurations with $M/M_{BH}$ in the interval $\left[ 0, 50 \right]$, whereas the plots on the right column correspond to configurations with $M/M_{BH}$ in the interval $\left[ 100, 1000 \right]$}.
\label{EHTconstraint-4}
\end{figure*}

The EHT collaboration imaged the central BH at M87 and Milk Way galaxies. The errors associated with these results were around $10\%$. This means that geometries deviating up to $10\%$ from the vacuum Schwarzschild/Kerr solutions cannot be distinguished with the current precision of the EHT. Based on this error of $10\%$, we can scan the ($M,a_0$)-plane and check whether a given hairy BH is favored or disfavored. In Fig.~\ref{EHT-constraint} we show the regions favored (green/lighter region) and disfavored (gray/darker region) by the current results of the EHT collaboration. These results were obtained by solving numerically Eqs.~\eqref{Eq-LR} and \eqref{crit-parameter} and computing the relative deviation with respect to the Schwarzschild result ($b_{crit}=3\sqrt{3}M_{\bh}$). If the relative deviation from the Schwarzschild result is less (greater) than $10\%$, the solution is in the favorable (disfavored) region.

Motivated by Ref.~\cite{Kocherlakota_etal:2021}, one can define another constraint in the parameters $M$ and $a_{0}$ by analyzing the shadow radius. The idea arises from prior knowledge of the angular gravitational radius of the M87* from stellar dynamics \cite{Gebhardt_etal:2011}. Hence, as presented in Ref.~\citep{M87_6:2019}, one can determine the deviation of the angular gravitational radius obtained by the 2017 EHT observations and the previously well-known value inferred from stellar dynamics. The difference found was $-0.01\pm 0.17$, for a $68\%$ level of confidence, which means that the inferred shadow size of M87*, as measured by the EHT collaboration, is consistent within a range of $17\%$ with the previous estimates, based on those stellar dynamics observations. This result can be translated into bounds on the allowed shadow radius $r_{\text{sh}}$, {\it i.e.,} 
	\begin{equation}
		4.31M_{\bh}\leq  r_{\text{sh}}\leq 6.08M_{\bh}.
	\label{ehtsradii}
	\end{equation}
In our case, due to spherical symmetry, $r_{\text{sh}}=b_{\text{crit}}$.%

Following an analogous rationale, one can use the recent SgrA* EHT data to constrain, even further, the values of the parameters $M$ and $a_0$. For the SgrA* case, the previous estimates on mass-to-distance ratio obtained with the analyses of stellar dynamics were predominantly produced by the Keck Observatory and the Very Large Telescope Interferometer (VLTI)~\cite{sgra_6:2022}. The corresponding bounds are:
	\begin{equation}\label{keckbound}
		4.5M_{\bh}\leq  r_{\text{sh}}\leq 5.5M_{\bh},
	\end{equation}
	for Keck and
	\begin{equation}\label{vltibound}
		4.3M_{\bh}\leq  r_{\text{sh}}\leq 5.3M_{\bh},
	\end{equation}
	for VLTI.
	  We exhibit in Fig.~\ref{EHTconstraint-4} the shadow radius of the hairy BH for different values of $M/M_\bh$ and $a_{0}/M_\bh$. In the top row of Fig.~\ref{EHTconstraint-4}, the green/lighter region denotes configurations compatible with the observations of the stellar dynamics for M87* [see Eq.~\eqref{ehtsradii}], while the gray/darker region denotes configurations disfavored by the observations of stellar dynamics.  In the bottom row of Fig.~\ref{EHTconstraint-4}, the blue and pink regions denote configurations compatible with the Keck and VLTI observations, respectively [see Eqs.~\eqref{keckbound} and \eqref{vltibound}]. The green region is the intersection of the two compatible regions (pink and blue). As in the top row of Fig.~\ref{EHTconstraint-4}, the gray region in the bottom row denotes configurations disfavored by the observations of stellar dynamic. We note that for fixed values of $a_0$, the shadow radius increases as we increase the halo mass $M$.  Also, some configurations compatible with the M87* bounds lie outside the region allowed by the SgrA* observations, for example, the cases with $a_0=1500M_\bh$ and $M/M_\bh \in [100,200]$.
	  
\section{Final remarks}\label{sec:remarks}
We have investigated light propagation in the spacetime of an asymptotically flat BH immersed in a dark matter halo described by a Hernquist profile, according to the solution reported in Ref.~\cite{Cardoso_etal:2021}. We have computed the LRs, shadows, and lensing for different values of the compactness parameter $\mathcal{C}$ and compared them with the Schwarzschild solution results.

Our study of the null geodesics has revealed that for a dilute dark matter halo (which corresponds to small values of $\mathcal{C}$) the position and height of the potential barrier are shifted, but without changing the overall behavior of the effective potential $H(r)$. For this case, the scattering of photons by the BH-dark matter system mimics what occurs in the vacuum case (Schwarzschild BH).  On the other hand, we found that more compact hairy BHs can support the existence of two unstable and one stable LR. Consequently, the scattering pattern changes significantly, as seen by the presence of more than one divergency in the scattering angle plot.

We have analyzed the gravitational lensing of the BH solution \eqref{Cardosoetalmetric} employing backward ray-tracing techniques. The parameters of this solution do not modify the spherical symmetry of the shadow. We concluded that for small values of the compactness parameter $\mathcal{C}$, the appearance of this hairy BH is similar to the Schwarzschild BH case. Higher values of $\mathcal{C}$ imply a larger shadow than in the Schwarzschild case. Moreover, the presence of two unstable and one stable LR in this regime leads to an intriguing lensing pattern.

We also performed a quantitative study of the shadows and showed that they can deviate from the Schwarzschild results by more than $10\%$. Using the EHT observations, we could rule out some halo configurations based on shadow radius bounds.  The majority of low-compactness configurations tested were compatible with the actual boundaries provided by the EHT results. Due to the high concentration of dark matter in the neighborhood of the BH, the systems with compactness greater than the unity yield shadow radii that lie outside the consistent regions of values for M87* and SgrA*.

Observational research on the strong gravity regime is currently experiencing an exciting era. Images of compact objects in galactic centers, such as those targeted by the EHT collaboration, provide an empirical evidence to explore the landscape of BH solutions.  In this context, evaluating the impact of the galaxy environment, as for instance, the effect of a dark matter halo, is crucial to constrain models for astrophysical BHs. A natural follow-up of our work would be to investigate the scenario with other light sources, for instance when the BH is illuminated by an accretion disk.

\begin{acknowledgments}

The authors would like to thank Carlos~A.~R.~Herdeiro and Pedro~V.~P.~Cunha for their important contributions to this work. We also thank the University of Aveiro, in Portugal, for the kind hospitality during the completion of this work. We acknowledge Funda\c{c}\~ao Amaz\^onia de Amparo a Estudos e Pesquisas (FAPESPA),  Conselho Nacional de Desenvolvimento Cient\'ifico e Tecnol\'ogico (CNPq) and Coordena\c{c}\~ao de Aperfei\c{c}oamento de Pessoal de N\'{\i}vel Superior (Capes) - Finance Code 001, in Brazil, for partial financial support. This work has further been supported by the European Union's Horizon 2020 research and innovation (RISE) programme H2020-MSCA-RISE-2017 Grant No. FunFiCO-777740 and by the European Horizon Europe staff exchange (SE) programme HORIZON-MSCA-2021-SE-01 Grant No. NewFunFiCO-101086251.
\end{acknowledgments}

	{}
\end{document}